\documentclass[preprint,12pt]{article}
\usepackage[dvips]{graphicx} %
\usepackage{pifont}
\usepackage{amsmath}
\usepackage{amsfonts}
\usepackage{amssymb}
\usepackage{color}
\usepackage{amsmath}
\usepackage{amssymb}
\usepackage{enumerate}
\usepackage{subfigure}
\usepackage{arydshln}
\usepackage{setspace}
\usepackage{multirow}
\usepackage{amssymb}
\usepackage{overpic}
\usepackage{booktabs}
\usepackage{amsmath}
\usepackage{txfonts}

\long
\def\symbolfootnote[#1]#2
{\begingroup
\def\thefootnote{\fnsymbol{footnote}}
\footnote[#1]{#2}\endgroup}
\usepackage{bm}

\begin{document}

\begin{center}
\textbf{A Spatial Structural Derivative Model for Ultraslow
Diffusion}
\end{center}

\begin{center}
Wei Xu$^{1}$, Wen Chen$^{1\ast }$, Yingjie Lian\textbf{g}$^{1\ast
}$, Jose Weberszpil$^{2}$
\end{center}

\begin{center}
$^{1}$ State Key Laboratory of Hydrology-Water Resources and
Hydraulic Engineering,\\
Institute of Soft Matter Mechanics, College
of Mechanics and Materials, Hohai University, Nanjing, China
\end{center}

\begin{center}
$^{2}$ Universidade Federal Rural do Rio de Janeiro, UFRRJ-IM/DTL
Av. Governador Roberto Silveira s/n- Nova Igua\c{c}\'{u}, Rio de
Janeiro, Brasil, 695014.\\

\end{center}
\begin{center}
 \textbf{Corresponding author:}
{chenwen@hhu.edu.cn, liangyj@hhu.edu.cn}
\end{center}
\noindent\textbf{Abstract: }This study investigates the ultraslow
diffusion by a spatial structural derivative, in which the
exponential function $e^{x}_{\, }$ is selected as the structural
function to construct the local structural derivative diffusion
equation model. The analytical solution of the diffusion equation is
a form of Biexponential distribution. Its corresponding mean squared
displacement is numerically calculated, and increases more slowly
than the logarithmic function of time. The local structural
derivative diffusion equation with the structural function $e^{x}$
in space is an alternative physical and mathematical modeling model
to characterize a kind of ultraslow diffusion.\\


\noindent\textbf{1. Introduction}\\

Anomalous diffusion [1, 2] has attracted great attention in diverse
fields, such as fractal porous media [3], polymer materials [4],
biomechanics [5], electrochemistry [6], and biomedical engineering
[7], just to mention a few. The mean squared displacement \,(MSD)
\,of anomalous function is a power law function of time [8, 9]:
\begin{equation}
\label{eq1}
\left\langle {x^{2}\left( t \right)} \right\rangle \propto t^{\eta },
\end{equation}
when $\eta >1$ characterizes super-diffusion, when $\eta <1$ is a
sub-diffusion, and it is a \,Brownian\, motion when $\eta =1$ [10].

Unlike the above-mentioned anomalous diffusion, ultraslow diffusion
also behaves in a dramatically different way from the normal
\,Brownian\, motion and is widely observed in nature and
engineering. It diffuses even far slower than the sub-diffusion
[11], such as the aging of high density colloids [12, 13], diffusion
of chemical solvents in polymerization [14], and atomic diffusion of
amorphous alloy melt [15]. The \,MSD\, of ultraslow diffusion is
often characterized by a logarithmic function of time in literature:
\begin{equation}
\label{eq2}
\left\langle {x^{2}\left( t \right)} \right\rangle \propto \left( {\ln t}
\right)^{\alpha },\;\;\;\;\alpha >0
\end{equation}
when $\alpha =4,_{\, }$ the \,MSD\, (2) reduces to the
classical Sinai diffusion law [16]. And when $\alpha =0.5$, it is
correlated with the well-known Harris law [17].

In order to provide more generalized description of ultraslow
diffusion, the structural derivative modeling methodology was
proposed [18], in which the structural function is chosen as the
inverse \,Mittag-Leffler\, function of time. Its corresponding
diffusion \,MSD\, is $\left\langle {x^{2}\left( t \right)}
\right\rangle \propto (E_{\alpha }^{-1} (t))^{\lambda }$, $\lambda
>0$, where $_{\, }E_{\alpha }^{-1} (t)$ is the inverse of \,Mittag-Leffler\,.
This novel expression generalizes the above ultraslow diffusion
including the logarithm ultraslow diffusion formulation (\ref{eq2})
[19] as its special case when the parameter $\alpha $ in the inverse
Mittag-Leffler function is 1. Instead of time inverse Mittag-Leffler
function ultraslow diffusion model, this study proposes a spatial
structural derivative ultraslow diffusion model via the structural
derivative in space, in which the exponential function $e^{x}_{\, }$
is chosen as the structural function. The analytical solution of the
diffusion model is derived by the scaling transform, and the
features of its \,MSD\, are further analyzed.

This paper is organized as follows: Section 2 introduces the spatial
structural derivative and proposes the exponential function
ultraslow diffusion model. In Section 3, the behaviors of normal and
sub-diffusions in comparison with the proposed diffusion models are
compared. Upon on the results and analysis of this study, the
conclusions are drawn in Section 4.\\

\noindent\textbf{2.} \textbf{Methodologies}\\

\noindent\textit{2.1 Structural derivative}\\

The structural derivative in space can be defined according to the
time structural derivative [20, 21]:
\begin{equation}
\label{eq3}
\frac{\mbox{d}p}{\mbox{d}_{s} x}=\lim\limits_{x_{1} \to x} \frac{p\left(
{x_{1} ,t} \right)-p\left( {x,t} \right)}{f\left( {x_{1} } \right)-f\left( x
\right)}\;
\end{equation}
where $S$ denotes the structural derivative, and $f\left( x \right)$
is the structural function. In Eq. (3), the structural
derivative is local and can be considered as a scaling transform:
\begin{equation}
\label{eq4}
\widehat{x}=f\left( x \right)
\end{equation}
when $f\left( x \right)=x$, Eq. (3) reduces to the classical
derivative in space [22], and when $f(x)=x^{\alpha }$, Eq.
(3) is the local fractal derivative [23].

The definition of the global structural derivative in space can be
derived from the global structural derivative in time [21],
\begin{equation}
\label{eq5}
\frac{\delta p(x,t)}{\delta_{s} x}=\frac{\partial }{\partial
x}\int\limits_{x_{1} }^x {k(x-\tau )} p(\tau ,t)d\tau
\end{equation}
which degenerates into the Riemann-Liouville fractional derivative
when $k(x)=\frac{x^{-\alpha }}{\Gamma (1-\alpha )}$ [24].

The classical derivative modeling strategy depicts the particular
factors on the rate of the change of time or space variables, but
less considers the important influence of the mesoscopic structure
of time-space fabric of the complex system on its physical
behaviors. While in the structural derivative, the structure
function depicts the time-space inherent property of the system,
which is a space-time transformation [25]. Consequently, the
structural derivatives can describe the causal relationship between
the mesoscopic space-time structure and the specific physical
quantity.\\

\noindent\textit{2.2 Spatial structural derivative equation model
for
ultraslow diffusion}\\

According to the local structural derivative, we establish the spatial
structural derivative model for ultraslow diffusion,
\begin{equation}
\label{eq6}
\frac{\mbox{d}p}{\mbox{d}t}=K\frac{\mbox{d}}{\mbox{d}_{s} x}\left(
{\frac{\mbox{d}p}{\mbox{d}_{s} x}} \right)
\end{equation}
where $K$ is the diffusion coefficient. When the structural function
$f\left( x \right)=x$, Eq. (6) yields a Gaussian
distribution [26]:
\begin{equation}
\label{eq7}
p\left( {x,t} \right)=\frac{1}{\sqrt {4\pi Kt} }\exp \left(
{-\frac{x^{2}}{4Kt}} \right)
\end{equation}

When $f\left( x \right)=x^{\beta }$, the solution of Eq. (6) is a stretched
Gaussian distribution:
\begin{equation}
\label{eq8}
p\left( {x,t} \right)=\frac{1}{\sqrt {4\pi Kt} }\exp \left(
{-\frac{x^{2\beta }}{4Kt}} \right)
\end{equation}

When $f\left( x \right)=e^{x},$ the corresponding structural
derivative is stated as:
\begin{equation}
\label{eq9}
\frac{\mbox{d}p}{\mbox{d}_{s} x}=\lim\limits_{x_{1} \to x} \frac{p\left(
{x_{1} ,t} \right)-p\left( {x,t} \right)}{e^{x_{1} }-e^{x}}\;
\end{equation}
and the corresponding solution of Eq. (6) can be derived
\begin{equation}
\label{eq10}
p\left( {x,t} \right)=\frac{1}{\sqrt {4\pi Kt} }\exp \left(
{-\frac{e^{2x}}{4Kt}} \right)
\end{equation}

Substituting the above formula (\ref{eq10}) into Eq. (\ref{eq6}) can easily verify
\begin{equation}
\label{eq11}
\frac{\mbox{d}}{\mbox{d}e^{x}}\left( {\frac{\mbox{d}p\left( {x,t}
\right)}{\mbox{d}e^{x}}} \right)=-p\left( {x,t} \right)\left(
{\frac{1}{2Kt}-\frac{e^{2x}}{4K^{2}t^{2}}} \right)=\frac{1}{K}\cdot
\frac{\mbox{d}}{\mbox{d}t}
\end{equation}

Namely, Eq. (10) is the solution of Eq. (6), in which the structural
function is an exponential function. Eq. (10) is a new kind of distribution,
called the Biexponential distribution in this paper. The relationship
between the structural function and the solution of structural derivative
diffusion equation in space is derived as:
\begin{equation}
\label{eq12}
p\left( {x,t} \right)=\frac{1}{\sqrt {4\pi Kt} }\cdot \exp \left(
{-\frac{\left( {f\left( x \right)} \right)^{2}}{4Kt}} \right)
\end{equation}

Generally speaking, the spatial structural derivative is a modeling strategy
and can be employed in modeling the ultraslow diffusion phenomena in complex
fluids. The solution of the corresponding structural derivative diffusion
equation constructed by the arbitrary structural function in the local
structural derivative in space is a kind of statistical distribution, i.e.,
the probability density function.

Fig. 1 is the probability density function described of Gaussian and
Biexponential distribution with $x>0,\;t=1,\;K=0.5$. From the
simulation results, we can see that the Biexponential distribution
decreases more rapidly than Gaussian distribution in a short time.
That means that compared with the probability of specify random
variables falling in a specific range, the Biexponential
distribution of tailing phenomenon is more evident.
\begin{figure}[htb]
\centering
\includegraphics[scale=0.8]{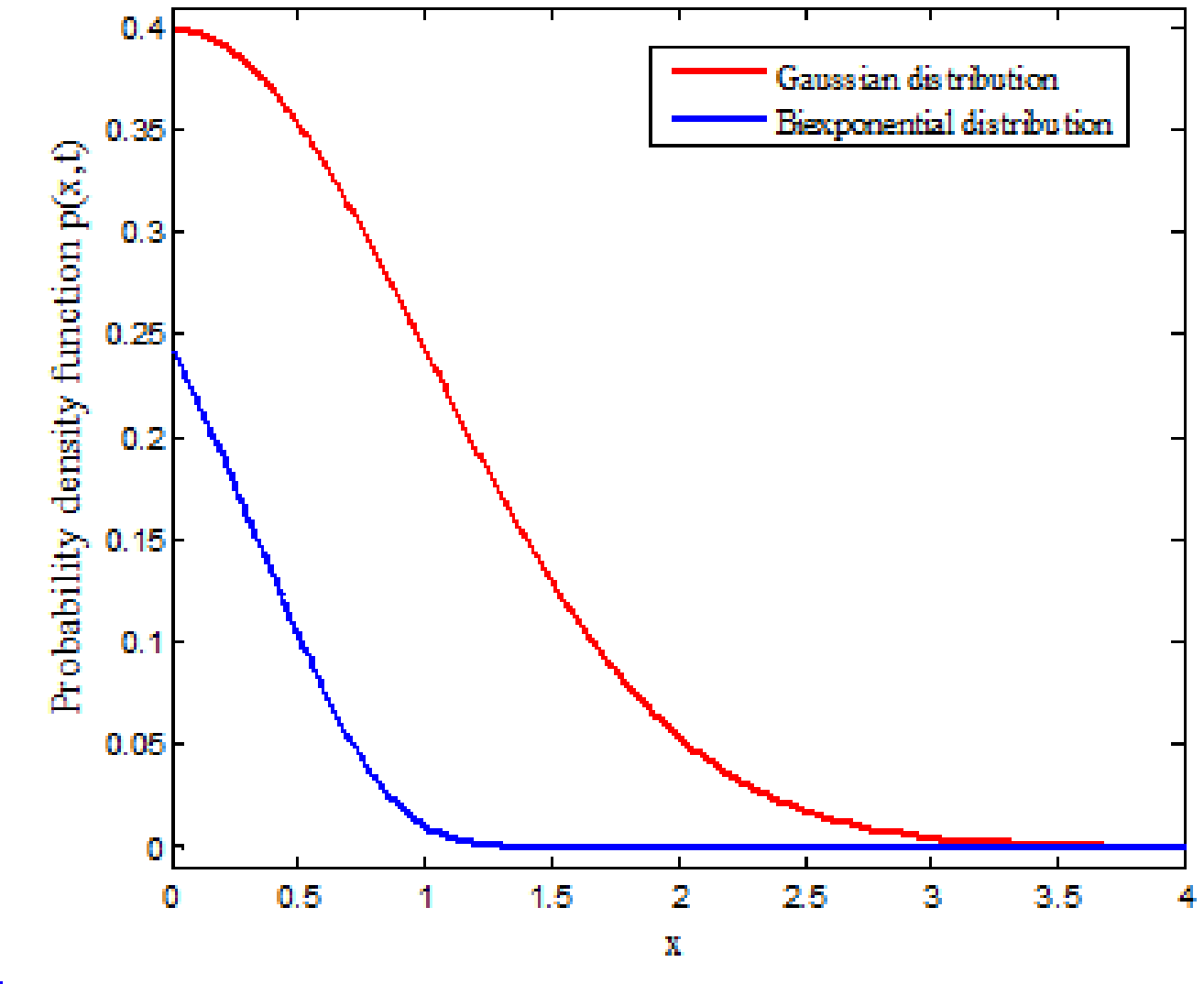}

{\textbf{Figure 1. The probability density function
 $t=1,\;K=0.5.$}}
\label{fig1}
\end{figure}

\noindent\textbf{3. }\textbf{Results and discussions}\\

In this section, we numerically compute the MSD of the proposed
ultraslow diffusion model, and then explore the transient diffusion
behavior by comparing with the normal diffusion, sub-diffusion,
super-diffusion, and the proposed ultraslow diffusions. Fig. 2 shows
the differences of various diffusion processes.\\

\begin{figure}[htb]
\centering
\includegraphics[scale=0.8]{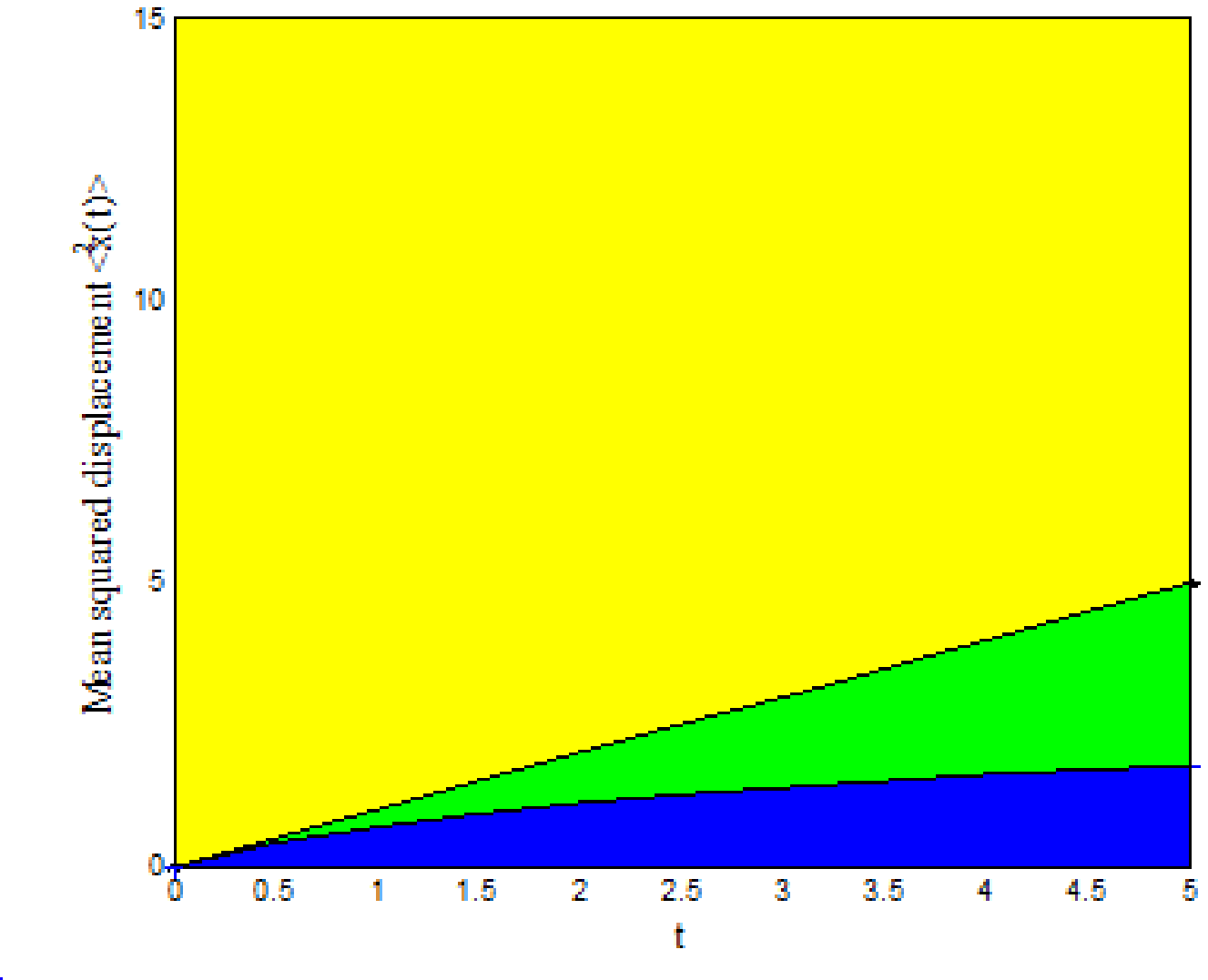}

\textbf{{Figure 2. Schematic diagram of normal diffusion,
sub-diffusion, super-diffusion, and the exponential structural
derivative ultraslow diffusion, in which the proposed ultraslow
diffusion and sub-diffusion is separated by the logarithm ultraslow
diffusion }$\left\langle {x^{2}\left( t \right)} \right\rangle =\ln
\left( {1+t} \right)$\textbf{ dotted with }$+$\textbf{, and the
normal diffusion }$\left\langle {x^{2}\left( t \right)}
\right\rangle =t$\textbf{ curve divides sub-diffusion and
super-diffusion dotted with *. }}
\label{fig1}
\end{figure}

In Fig. 2, the yellow area represents the super-diffusion process, the
corresponding MSD is $\left\langle {x^{2}\left( t \right)} \right\rangle
=\left( {t\mbox{+}1} \right)^{\beta },\;\beta >1$. The blue and the green
areas respectively belong to the ultraslow diffusion and sub-diffusion.

The MSD of the proposed exponential function ultraslow diffusion can be
derived from Eq. (\ref{eq10}) as
\begin{equation}
\label{eq13}
\left\langle {x^{2}(t)} \right\rangle =\int\limits_{\mbox{-}\infty }^\infty
{x^{2}} p\left( {x,t} \right)\mbox{d}x=\frac{1}{\sqrt {4Kt\pi }
}\int\limits_{\mbox{-}\infty }^\infty {x^{2}} \cdot \exp \left(
{-\frac{e^{2x}}{4Kt}} \right)\mbox{d}x
\end{equation}

Its analytical solution can not directly be obtained, instead we define the
MSD in $\left( {0,\;+\infty } \right)$ and calculate the following integral
form
\begin{equation}
\label{eq14}
\left\langle {x^{2}(t)} \right\rangle =\int\limits_0^\infty {x^{2}} p\left(
{x,t} \right)\mbox{d}x=\frac{1}{\sqrt {4Kt\pi } }\int\limits_0^\infty
{x^{2}} \cdot \exp \left( {-\frac{e^{2x}}{4Kt}} \right)\mbox{d}x
\end{equation}\\

\begin{figure}[htb]
\centerline{\includegraphics[width=5.69in,height=4.61in]{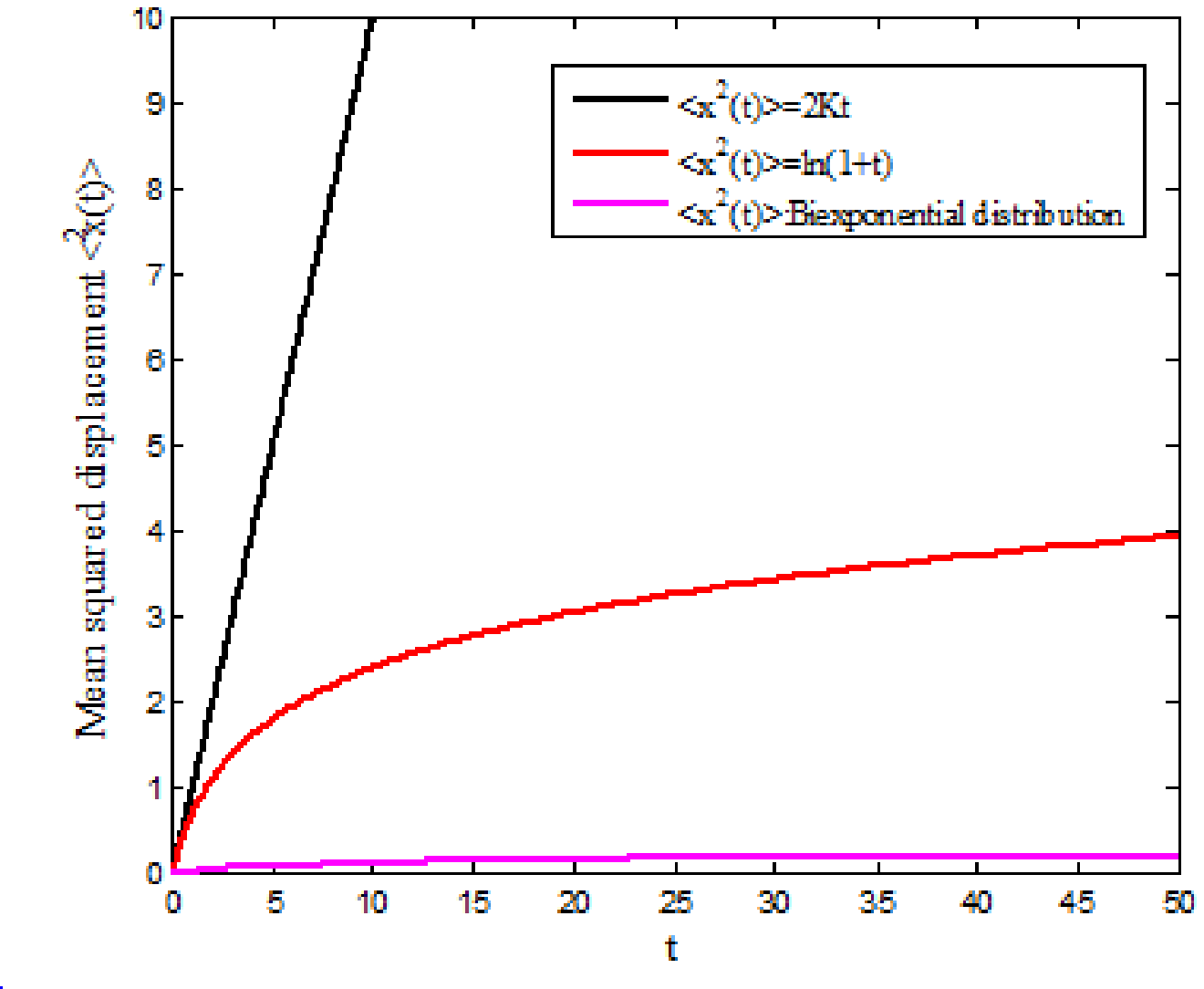}}

\textbf{Figure 3. Mean squared displacement of normal diffusion,
logarithm ultraslow diffusion and exponential function ultraslow
diffusion with}$K=0.5$\textbf{. }
\label{fig3}
\end{figure}

Fig. 3 shows the MSD of normal diffusion, logarithm ultraslow diffusion and
the present exponential structural function ultraslow diffusion. We can
observe from Fig.3 that the MSD of the proposed ultraslow diffusion
increases slower with time than that of the logarithmic diffusion. Thus the
local structural derivative diffusion equation with the structural function
$f\left( x \right)=e^{x}$ in space is a mathematical modeling
method to characterize a kind of ultraslow diffusion.

It is worthy of noting that the exponential function$f\left( x
\right)=e^{x}$ is a special case of the popular Mittage-Leffler function,
\begin{equation}
\label{eq15}
E_{\alpha } \left( x \right)=\sum\limits_{k=0}^\infty {\frac{x^{k}}{\Gamma
\left( {\alpha k+1} \right)}}
\end{equation}
when $\alpha =1$, it degenerates into the exponential function.

In recent years, the Mittage-Leffler function has widely been used
in the fractal dynamics, anomalous diffusion and fractal random
field [27-29]. In addition, the inverse Mittage-Leffler function as
has also been applied to describe ultraslow diffusion [25]. In
further study, we will try to investigate different structural
functions with clear physical mechanism, such as Mittag-Leffler
function and its inverse function, to construct both local and
global structural derivative diffusion equation in modeling
non-Gaussian motion.\\

\noindent4. \textbf{Conclusions}\\

In this paper, we present a local spatial structural derivative diffusion
model to depict the ultraslow diffusion, in which the exponential function
$\mbox{e}^{x}$ is selected as the structural function. Based on the
foregoing results and discussions, the following conclusions can be drawn:

1. The analytical solution of the proposed ultraslow diffusion
equation is a form of Biexponential distribution.

2. The corresponding mean squared displacement is numerically calculated,
and increases more slowly with than that of the logarithmic ultraslow
diffusion.

3. The local structural derivative diffusion equation with the
structural function $e^{x}$ in space is an alternative mathematical
modeling method to characterize a kind of ultraslow diffusion.

\noindent\textbf{Acknowledgment}\\

This paper was supported by the National Science Funds for
Distinguished Young Scholars of China (Grant No. 11125208) and the
111 project (Grant No. B12032).\\


\begin{thebibliography}{6}

\bibitem{} R. Gorenflo, F. Mainardi, D. Moretti, G. Pagnini, P. Paradisi, Discrete random walk models for space-time fractional diffusion, \textit{Chemical Physics}, \textit{284} (2002), 1-2, pp. 521-541.

\bibitem{} R. Metzler, J. Klafter, The random walk's guide to anomalous diffusion: a fractional dynamics approach, \textit{Physics Reports}, \textit{339} (2000), 1, pp. 1-77.

\bibitem{}  Z. Yong, D.A. Benson, M.M. Meerschaert, H.P. Scheffler, On using random walks to solve the space-fractional advection-dispersion equations, \textit{Journal of Statistical Physics}, \textit{123 }(2006), 1, pp. 89-110.

\bibitem{}  F. Mainardi, Fractional calculus and waves in linear viscoelasticity, Imperial College Press, World Scientific, 2010.

\bibitem{}  C. Ingo, R.L. Magin, L. Colon-Perez, W. Triplett, T.H. Mareci, On random walks and entropy in diffusion-weighted magnetic resonance imaging studies of neural tissue, \textit{Magnetic Resonance in Medicine}, \textit{71 }(2014), 2, pp. 617-627.

\bibitem{}  A. Ehsani, M.G. Mahjani, M. Bordbar, S. Adeli, Electrochemical study of anomalous diffusion and fractal dimension in poly ortho aminophenol electroactive film: Comparative study, \textit{Journal of Electroanalytical Chemistry}, \textit{710 }(2013), 12, pp. 29-35.

\bibitem{}  M. K�pf, C. Corinth, O. Haferkamp, T.F. Nonnenmacher, Anomalous diffusion of water in biological tissues, \textit{Biophysical Journal}, \textit{70 }(1996), 6, pp. 2950-2958.

\bibitem{}  M.F. Shlesinger, Asymptotic solutions of continuous-time random walks, Journal of Statistical Physics, \textit{10} (1974), 5, 421-434.

\bibitem{}  G.H. Weiss, R.J. Rubin, Random Walks: Theory and Selected Applications, 1983.

\bibitem{}  R. Metzler, J.H. Jeon, A.G. Cherstvy, E. Barkai, Anomalous diffusion models and their properties: non-stationarity, non-ergodicity, and ageing at the centenary of single particle tracking, \textit{Physical Chemistry Chemical Physics}, \textit{16}(2014), 44,pp. 24128-24164.

\bibitem{}  S. Hou, Stochastic model for ultraslow diffusion, \textit{Stochastic Processes {\&} Their Applications}, \textit{116} (2006), 116, pp. 1215-1235.

\bibitem{}  S. Boettcher, P. Sibani, Ageing in dense colloids as diffusion in the logarithm of time, \textit{J Phys Condens Matter}, \textit{23} (2011), 6, pp. 65-103.

\bibitem{}  R.E. Courtland, E.R. Weeks, Direct visualization of aging in colloidal glasses, \textit{Journal of Physics Condensed Matter}, \textit{15} (2002), 1, pp. S359.

\bibitem{}  M.L. Longmire, M. Watanabe, H. Zhang, T.T. Wooster, R.W. Murray, Voltammetric measurement of ultraslow diffusion rates in polymeric media with microdisk electrodes, \textit{Analytical Chemistry}, \textit{62} (1990), 7, pp. 747-752.

\bibitem{}  H. Jinliang, Z. Chunao, G. Yongliang, Z. Bo, Atomic diffusion in amorphous alloy melts, \textit{Materials China}, \textit{5} (2014) pp. 282-288.

\bibitem{}  Y.G. Sinai, The limiting behavior of a one-dimensional random walk in a random medium, \textit{Theory of Probability {\&} Its Applications}, \textit{27} (2006), 2, pp. 256-268.

\bibitem{}  M.A. Lomholt, L. Lizana, R. Metzler, T. Ambj�rnsson, Microscopic origin of the logarithmic time evolution of aging processes in complex systems, \textit{Physical Review Letters}, \textit{110 }(2013), 20, pp. 208301.

\bibitem{}  W. Chen, Y. Liang, X. Hei, Local structure derivative and its application, \textit{Journal of Solid Mechanics}, \textit{37} (2016), 5, pp. 456-460.

\bibitem{}  W. Chen, Y. Liang, X. Hei, Structural derivative based on inverse Mittag-Leffler function for modeling ultraslow diffusion, \textit{Fractional Calculus {\&} Applied Analysis}, \textit{19} (2016), 5, pp. 1316-1346.

\bibitem{}  W. Chen, Time-space fabric underlying anomalous diffusion, \textit{Chaos Solitons {\&} Fractals},\textit{ 28 }(2006), 4,
pp. 923-929.

\bibitem{}  W. Chen, X. Hei, Y. Liang, A fractional structural derivative model for ultra-slow diffusion, \textit{Applied Mathematics and Mechanics}, \textit{37} (2016), 6, pp. 599-608.

\bibitem{}  W. Chen, H. Sun, X. Li, Fractional Derivative Modeling of Mechanics and Engineering Problems, \textit{Science Press}, Beijing, 2010.

\bibitem{}  X.J. Yang, H.M. Srivastava, J.H. He, D. Baleanu, Cantor-type cylindrical-coordinate method for differential equations with local fractional derivatives, \textit{Physics Letters A}, \textit{377} (2013), 28-30, pp. 1696-1700.

\bibitem{}  I. Podlubny, Fractional Differential Equations, Academic press, 1999.

\bibitem{}  Wen Chen, Yingjie Liang, Xindong Hei, Structural derivative based on inverse Mittag-Leffler function for modeling ultraslow diffusion, \textit{Fractional Calculus {\&} Applied Analysis}, \textit{19}(2016), 5, pp.1316-1346.

\bibitem{}  Y. Bazi, L. Bruzzone, F. Melgani, Image thresholding based on the EM algorithm and the generalized Gaussian distribution, \textit{Pattern Recognition}, \textit{40} (2007), 2, pp. 619-634.

\bibitem{}  Chaurasia, B.L. V., Pandey, C. S., On the fractional calculus of generalized Mittag-Leffler function, \textit{Sci.ser.a Math.sci}, 20 (2010), pp. 113-122.

\bibitem{}  E.C.D. Oliveira, F. Mainardi, J.V. Jr, Models based on Mittag-Leffler functions for anomalous relaxation in dielectrics, \textit{The European Physical Journal Special Topics}, \textit{193} (2011), 1, pp. 161-171.

\bibitem{} C. Zeng, Y.Q. Chen, Global Pad'e approximations of the generalized Mittag-Leffler function and its inverse, \textit{Fractional Calculus {\&} Applied Analysis}, \textit{18} (2013), 8, pp. 1492-1506.

\end{thebibliography}
\end{document}